\documentclass{moriond}

\def\be{\begin{equation}}
\def\ee{\end{equation}}
\def\bea{\begin{eqnarray}}
\def\eea{\end{eqnarray}}

\newcommand{\Photo}{}
\def\lb{\textit{LiteBIRD}}

\def\planck{\textit{Planck}}
\usepackage{amsmath} 
\usepackage{amssymb}

\begin{document}
\vspace*{4cm}
\title{Contribution to the 2022 Cosmology session of the 56th Rencontres de Moriond: Moment expansion of polarized dust SED: A new path towards capturing the CMB $B$-modes with \lb{}}

\author{L. Vacher, J. Aumont, L. Montier, S. Azzoni, F. Boulanger and M. Remazeilles for the \lb\ Collaboration }

\address{IRAP, Universit\'e de Toulouse, CNRS, CNES, UPS,\\ Toulouse, France}

\maketitle\abstracts{
Characterizing accurately the polarized dust emission from our Galaxy will be decisive for the quest for the Cosmic Microwave Background (CMB) primordial $B$-modes. The incomplete modeling of its potentially complex spectral properties could lead to biases in the CMB polarization analyses and to a spurious detection of the tensor-to-scalar ratio $r$. Variations of the dust properties along and between lines of sight lead to unavoidable distortions of the spectral energy distribution (SED) that can not be easily anticipated by standard component separation methods. This issue can be tackled using a moment expansion of the dust SED, an innovative parametrization method imposing minimal assumptions on the sky complexity.
In the recent work [Vacher \emph{et al.} (2022)]\cite{Vacher_2022}, we apply this formalism to the $B$-mode cross-angular power spectra computed from simulated \lb{} polarization data at frequencies between 100 and 402\,GHz, containing CMB, dust and instrumental noise. Thanks to the moment expansion, we can measure an unbiased value of the tensor-to-scalar ratio with a dispersion compatible with the target values aimed by the instrument.
}

\section{Introduction \label{sec:introduction}}

Several astrophysical sources emit polarized light in the same frequency range as the cosmic microwave background (CMB) fluctuations, with an intensity that can be several orders of magnitudes greater. Two main contributions are expected to be significant: at low frequencies ($\leq 70$\,GHz), the synchrotron emission dominates. It is due to charged light particles, often coming from ionized regions, accelerated circularly in the Galactic magnetic field. At higher frequencies, thermal dust signal is the main contribution. Dust grains are forged in the envelopes of pulsing massive stars and supernovae from which they are expelled into the interstellar medium where they will play a key role in the Galactic dynamics and chemistry. Dust grains are heated by starlight which they re-emit in microwave and infrared. At least $\sim 30\%$ of the starlight of the universe is reradiated that way [Bernstein \emph{et al.} (2002)]\cite{Bernstein_2002}. Because of their elongated shape, the dust grains will have a preferred alignment in the galactic magnetic field and their signal will be strongly polarized. As such, dust grains will create strong $B$-modes signal over the sky, largely exceeding the predicted primordial ones. 

In order to seek for the faint leftover signal from inflation in the CMB, it is thus critical to identify the dust contribution to the total $B$-mode emission in order to separate both signals. Doing so is highly non trivial and can lead to spurious measurement of the tenso-to-scalar ratio $r$, quantifying the intensity of the primordial $B$-modes.

The canonical way to model the frequency dependence of the dust signal -- its spectral energy distribution (SED) -- in a given line of sight $\vec{n}$, is given by the modified black body (MBB):
\begin{equation}
\label{eq:MBB}
     I(\nu,\vec{n}) =  \left(\frac{\nu}{\nu_0}\right)^{\beta(\vec{n})} \frac{B_{\nu}(T(\vec{n}))}{B_{\nu_0}(T(\vec{n}))} A(\vec{n}) 
     =\frac{I_{\nu}(\beta(\vec{n}),T(\vec{n}))}{I_{\nu_0}(\beta(\vec{n}),T(\vec{n}))} A(\vec{n}),
\end{equation}
\noindent which is a black body function $B_\nu$ at a temperature $T_0$ multiplied by the frequency $\nu$ to the power of the spectral index $\beta_0$. $A$ is the amplitude of the dust signal across the sky. The overall SED is normalized by a MBB with a reference frequency $\nu_0$. 
The MBB function is an empirical model that has proven to provide a robust description of the dust signal. However, it is non linear: the sum of two different MBBs do not result in a MBB. The MBB is then not a good model to fit over mixed MBB signals. This is problematic since, in true experimental conditions, averages of SED coming from different sky regions can not be avoided: along the line of sight; between different lines of sight, inside the beam of the instrument or; when doing a spherical harmonic decomposition to calculate the angular power spectra over large regions of the sky. These averages will deform the SED away from its canonical model, these deformations are called SED distortions. Miss-modelling those distortions can easily lead to confuse dust $B$ modes and primordial ones, leading to a spurious measurement of the tensor-to-scalar ratio.

\section{The moment expansion formalism \label{sec:formalism}}

The moment expansion, proposed in [Chluba \emph{et al.} (2017)]\cite{Chluba_2017} aims to model these SED distortions with a Taylor inspired expansion of the SED with respect to its spectral parameters. For the MBB, the expansion is done with respect to $\beta$ and $T$ around the pivot values $\beta_0$ and $T_0$: 

\begin{align}
     I(\nu,\vec{n})  = \frac{I_{\nu}(\beta_0,T_0)}{I_{\nu_0}(\beta_0,T_0)} \bigg\{ & A(\vec{n}) + \omega^\beta_1(\vec{n}) \ln\left(\frac{\nu}{\nu_0}\right)+ \frac{1}{2}\omega^\beta_2(\vec{n}) \ln^2\left(\frac{\nu}{\nu_0}\right)\nonumber \\[2mm]
     &+ \omega^T_1(\vec{n})\Big( \Theta(\nu,T_0) - \Theta(\nu_0,T_0) \Big) + \dots \bigg\},
\label{eq:momentinttemp}
\end{align}

\noindent where $\Theta$ is the derivative of $B_\nu$ with respect to $T$. The coefficients $\omega^{p}_i$ are the so called \emph{moments} of order $i$ with respect to the parameter $p$, that quantify the amplitude of the SED distortions.
This expression can be generalized at the cross-frequency power spectra level as in [Mangilli \emph{et al.} (2021)]\cite{Mangilli_2021}:

\begin{align}
    \mathcal{D}_\ell(\nu_i \times \nu_j) &= \frac{I_{\nu_i}(\beta_0(\ell),T_0(\ell))I_{\nu_j}(\beta_0(\ell),T_0(\ell))}{I_{\nu_0}(\beta_0(\ell),T_0(\ell))^2} \cdot \bigg\{ \nonumber \\[-0.5mm]
    0^{\rm th}\ \text{order}\;&
    \begin{cases}
    & \ \mathcal{D}_\ell^{A \times A}
    \end{cases}
    \nonumber \\[-0.5mm]
    1^{\rm st}\ \text{order}\ \beta\;&
    \begin{cases}
    &+\mathcal{D}_\ell^{A \times \omega^{\beta}_1}\left[ \ln\left(\frac{\nu_i}{\nu_0}\right) + \ln\left(\frac{\nu_j}{\nu_0}\right) \right] \nonumber \\
    &+ \mathcal{D}_\ell^{\omega^{\beta}_1 \times \omega^{\beta}_1} \left[\ln\left(\frac{\nu_i}{\nu_0}\right)\ln\left(\frac{\nu_j}{\nu_0}\right) \right]\nonumber \\  \end{cases}\\[-0.5mm]
    1^{\rm st}\ \text{order}\ T \;&
    \begin{cases}
    &+\mathcal{D}_\ell^{A \times \omega_1^T} \left( \Theta_i +  \Theta_j -  2\Theta_0\right) \\
    &+ \mathcal{D}_\ell^{\omega_1^T \times \omega_1^T}\Big(\Theta_i - \Theta_0\Big)\left(\Theta_j - \Theta_0\right)\nonumber
    \end{cases}\\[-0.5mm]
    1^{\rm st}\ \text{order}\ T\beta \;&
    \begin{cases}
    &+ \mathcal{D}_\ell^{\omega^{\beta}_1 \times \omega_1^T} \left[ \ln\left(\frac{\nu_j}{\nu_0} \right)\Big( \Theta_i - \Theta_0\Big) + \ln\left(\frac{\nu_i}{\nu_0} \right)\left( \Theta_j - \Theta_0\right)\right]  \nonumber \\
    \end{cases}\\[-0.5mm]
    2^{\rm nd}\ \text{order}\ \beta \;&
    \begin{cases}
    &+ \frac{1}{2} \mathcal{D}_{\ell}^{A \times\omega^{\beta}_{2}}  \left[ \ln^2\left(\frac{\nu_i}{\nu_0}\right)
    + \ln^2\left(\frac{\nu_j}{\nu_0}\right)
    \right]
    \\[-0.5mm]
    &+ \frac{1}{2} \mathcal{D}_{\ell}^{\omega^{\beta}_{1} \times \omega^{\beta}_{2}}  \Big[ \ln \left(\frac{\nu_i}{\nu_0}\right)  \ln^2\left(\frac{\nu_j}{\nu_0}\right)  
    +\ln\left(\frac{\nu_j}{\nu_0}\right)
    \ln^2 \left(\frac{\nu_i}{\nu_0}\right) \Big] 
    \\[-0.5mm]
    &+\frac{1}{4} \mathcal{D}_{\ell}^{\omega^{\beta}_{2} \times \omega^{\beta}_{2}}  \left[\ln^2 \left(\frac{\nu_i}{\nu_0}\right)  \ln^2 \left(\frac{\nu_j}{\nu_0}\right) \right]\quad+ \dots \bigg\},
    \,
    \end{cases}\nonumber\\[-0.5mm]
    &
\label{eq:moments}
\end{align}

\noindent where $\mathcal{D}_\ell(\nu_i \times \nu_j)= \frac{\ell(\ell+1)}{2\pi}\mathcal{C}_\ell({\rm map}(\nu_i)\times{\rm map}(\nu_j))$.
This expression provides an analytical expression that can model dust SED with varying spectral parameters over the sky. The $\mathcal{D}^{A\times B}_\ell$ are the free parameters to estimate, quantifying the SED distortions. To define various fitting schemes, we cut the above expansion at different orders: up to order 0 (MBB), up to order 1 in $\beta$ ($\beta$-1), up to order 1 in both $\beta$ and $T$ ($\beta$-$T$) and up to order 2 for $\beta$ only ($\beta$-2).

\section{Application to \lb \label{sec:results}}

\begin{figure}
\centering
\begin{minipage}{0.35\linewidth}
\centerline{\includegraphics[scale=0.35]{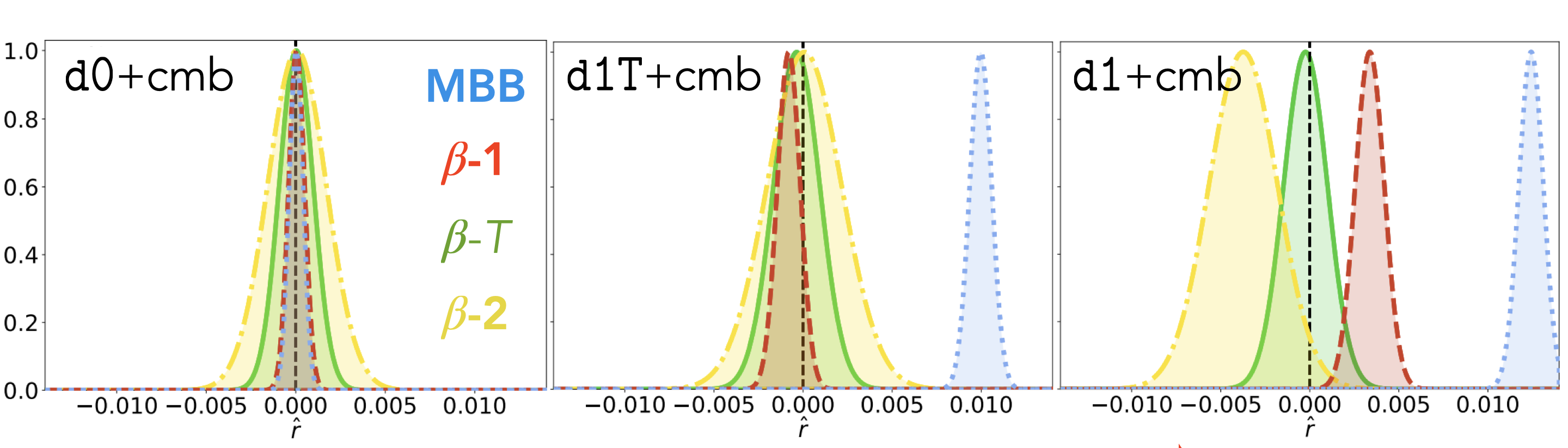}}
\end{minipage}
\caption[]{Recovered posterior for the tensor to scalar ratio with the three different kind of dust models; {\tt d0} (left), {\tt d1T} (center) and {\tt d1} (right) with the different fitting schemes: MBB (blue), $\beta$-1 (red), $\beta$-$T$ (green) and $\beta$-2 (yellow). The black dashed line indicates $r_{\rm sim}=0$.}
\label{fig:posteriors}
\end{figure}

In order to test this method with \lb{}, we generate $(I,Q,U)$ simulated maps including Gaussian instrumental noise at the 9 highest frequency bands of the instrument ($\ge 100$\,GHz). We mask the maps in order to keep a large sky fraction of $f_{\rm sky}=0.7$. Three different dust models are considered, containing a MBB of increasing complexity in every pixel: {\tt d0} in which both $\beta$ and $T$ are constant over the sky, including no SED distortions, {\tt d1T} having $T$ constant but a spatially varying spectral index $\beta(\vec{n})$ and {\tt d1} where both $\beta$ and $T$ are allowed to vary spatially. The amplitude and varying spectral parameter templates are taken from the \planck{} 2015 data at 353\,GHz and extrapolated to a frequency $\nu$ using the corresponding MBB. 
We generate $N_{\rm sim}=500$ simulations with each dust type. For every simulation, we add a Gaussian contribution of CMB with $r_{\rm sim}=0$. 

We then calculate the cross-frequency power spectra $\mathcal{D}^{\rm sim}_\ell(\nu_i \times \nu_j)$ for every simulation. Keeping 9 bands, we end up with 45 cross-frequency spectra. Only the $B\times B$ auto spectra are consider in the analysis.
For every simulation, the following model is fitted over the extracted cross-frequency power spectra:

\begin{equation}
  \mathcal{D}_{\ell}^{\rm model}(\nu_i \times \nu_j) = \mathcal{D}_{\ell}^{{\rm dust}}\left(\beta_0(\ell), T_0(\ell), \mathcal{D}^{\mathcal{M}\times\mathcal{N}}_{\ell}(\nu_i\times\nu_j)\right) +  \mathcal{D}_{\ell}^{{\rm lensing}} + r \cdot  \mathcal{D}_{\ell}^{{\rm tensor}}, 
\label{eq:model}
\end{equation}

\noindent where $\mathcal{D}_{\ell}^{{\rm dust}}$ is given by Eq.~\ref{eq:moments} with the various fitting schemes described above. $\mathcal{D}_{\ell}^{{\rm lensing}}$ is the CMB lensed $E$ modes contribution to the $B$-modes in the simulation and $\mathcal{D}_{\ell}^{{\rm tensor}}$ is the expected theoretical primordial $B$ mode spectra. 

The following $\chi^2$ is minimized:

\begin{equation}
\chi^2 = (\mathcal{D}^{\rm sim}_\ell - \mathcal{D}^{\rm model}_\ell)^T\mathbb{C}^{-1}(\mathcal{D}^{\rm sim}_\ell - \mathcal{D}^{\rm model}_\ell),
\label{eq:chi2}
\end{equation}

\noindent where $\mathbb{C}$ is the covariance given by $\mathbb{C}_{\ell,\ell'}^{i\times j,k\times l} = {\rm cov}\left(\mathcal{D}^{\rm sim}_\ell (\nu_i \times \nu_j),\mathcal{D}^{\rm sim}_{\ell'} (\nu_k \times \nu_l)\right).$
 
After $\chi^2$ minimization, it can be shown that the moments are significantly detected only when the dust content is more complex than ${\tt d0}$. They thus remain compatible with zero if there is no SED distortions in the signal but are used by the fit when such distortions are present. 
A single best-fit value of the tensor-to-scalar ratio $\hat{r}$ is obtained for each simulation. An histogram can be built with the $N_{\rm sim}$ values of $\hat{r}$ over which we can fit a Gaussian curve as displayed in Fig.~\ref{fig:posteriors}. The fitted Gaussian standard deviation is noted $\sigma_{\hat{r}}$. For the simplest case {\tt d0}, all fitting schemes allows to recover a value of $\hat{r}$ centered on $r_{\rm sim}=0$, as desired. For {\tt d1T} and {\tt d1}, the MBB is not a good fit anymore due to the presence of the SED distortions mentioned above. The posteriors are strongly biased ($\sim 20 \sigma_{\hat{r}}$). Adding some moments with the $\beta$-1 fitting scheme allows to reduce the bias of the posterior for {\tt d1T} and {\tt d1} but is not enough to get a posterior compatible with $r_{\rm sim}$ at $1 \cdot \sigma_{\hat{r}}$. The $\beta$-$T$ fitting scheme centers the posteriors for both {\tt d1T} and  {\tt d1} with a width of $\sim 1\times 10^{-3}$, providing a good model for the SED distortions in both cases. While the $\beta$-2 fitting scheme correctly model the spectral distortions in {\tt d1T}, an unexpected negative bias appears for {\tt d1}. As discussed in [Vacher \emph{et al.} (2022)]\cite{Vacher_2022}, this bias is due to the failure of the $\beta$ moments to model the temperature distortions and a strong degeneracy between these moments and the CMB signal.

\section{Conclusion and discussion\label{sec:conclusion}}

Moment expansion in harmonic space provides a powerful tool to model the SED distortions coming from averaging effects of spatially varying spectral parameters. Such a modeling will be necessary for next generations of CMB experiments as \lb{}. Using the first order moments in both $\beta$ and $T$ allows to recover an unbiased value of the tensor-to-scalar ratio $r$ in all the scenarios considered, even the most complex one where the dust signal is given by a MBB in every pixel, with spatially varying spectral index and temperature over the sky. In [Vacher \emph{et al.} (2022)]\cite{Vacher_2022}, it has been shown that this result is robust using different sky fractions, adding a synchrotron component and when adding a non zero value of $r_{\rm sim}$. The method can also be optimized in order to keep only the necessary coefficients in order to decrease the recovered value of $\sigma_{\hat{r}}$. However the correction provided by the moments seems to be strongly dependent of the moment frequency dependence themselves and expending around the wrong canonical model could have strong implications. Moreover, some correlations between some of these moments and the CMB could be a strong limitation in more complex scenarios than the ones considered here.

\end{document}